\documentclass[12pt,preprint]{aastex}
\usepackage{graphicx}
\shorttitle{The Old Nova V842 Cen} 
\shortauthors{Sion et al.} 

\begin{document}

\title{Multiwavelength Photometry and Hubble Space Telescope Spectroscopy of the Old Nova V842 Centaurus\altaffilmark{1}}

\author{Edward M. Sion} 
\affil{Department of Astronomy \& Astrophysics, Villanova University, \\ 
800 Lancaster Avenue, Villanova, PA 19085, USA}
\email{edward.sion@villanova.edu}
 
\author{Paula Szkody, Anjum Mukadam}
\affil{Department of Astronomy, University of Washington, \\ Seattle, WA 98195, USA}
\email{szkody@astro.washington.edu, anjum@astro.washington.edu}

\author{Brian Warner, Patrick Woudt}
\affil{Astrophysics, Cosmology and Gravity Centre, Department of Astronomy, University of Cape Town, \\ Private Bag X3, Rondebosch 7701, South Africa}
\email{brian.warner@uct.ac.za, pwoudt@ast.uct.ac.za}

\author{Frederic Walter}
\affil{Dept. of Physics and Astronomy, Stony Brook University, \\ Stony Brook, NY 11794, USA}
\email{frederick.walter@stonybrook.edu}

\author{Arne Henden}
\affil{AAVSO \\ 49 Bay State Road, Cambridge, MA 02138, USA}
\email{arne@aavso.edu}

\and

\author{Patrick Godon}
\affil{Depatment of Astronomy \& Astrophysics, \\ Villanova University, Villanova, PA 19085, USA}
\email{patrick.godon@villanova.edu}

\altaffiltext{1}{\it Based on observations made with the NASA/ESA Hubble Space Telescope, obtained at the Space
Telescope Science Institute, which is operated by the Association of Universities for Research in Astronomy,
Inc., (AURA) under NASA contract NAS 5-26555, with the Apache Point Observatory 3.5m telescope which
is owned and operated by the Astrophysical Research Consortium.}

\begin{abstract} 

We present ground-based optical and near infrared photometric
observations and Hubble Space Telescope COS spectroscopic
observations of the old nova V842 Cen (Nova Cen 1986).
Analysis of the optical light curves reveals a peak at 
56.5$\pm$0.3s with an amplitude of 8.9$\pm$4.2 mma, which is consistent with the rotation
of a magnetic white dwarf primary in V842 Cen that was detected earlier 
by Woudt et al., and led to its classification as an
intermediate polar.However, our UV lightcurve
created from the COS time-tag spectra does not show this periodicity.
Our synthetic spectral analysis of an HST COS spectrum
rules out a hot white dwarf photosphere as the source of the FUV flux.
The best-fitting model to the COS spectrum is a full optically thick accretion disk
with no magnetic truncation, a low disk inclination angle, low accretion rate and 
a distance less than half the published distance that was determined on the basis 
of interstellar sodium D line strengths.Truncated accretion disks with truncation 
radii of 3$R_{wd}$ and 5$R_{wd}$ yielded unsatisfactory agreement with the COS data. 
The accretion rate is unexpectedly low
for a classical nova only 24 years after the explosion when the accretion rate 
is expected to be high and the white dwarf should still be very hot, especially 
if irradiation of the donor star took place.
Our low accretion rate is consistent with low accretion rates derived from X-ray and ground-based
optical data. 
\end{abstract}
\section{Introduction}          
The old nova V842 Cen (Nova Cen 1986) was discovered on 1986 November 22 by 
McNaught (1986) and reached a maximum magnitude of 5.6 from its pre-outburst
brightness of about 18.5. Whitelock (1987) and Sekiguchi et al. (1989) estimated
the $t_3$ decay time to be 48 d, thus classifying it as a moderately
fast nova. Since 1987, this object has been followed with 
photometric and spectroscopic observations from the ground and in space. A 
summary of observations through 2005 is given by Schmidtobreick et al. (2005) who concluded at that time that there was still insufficient time for the nova to have cooled down as either the white dwarf itself or a region of the accretion disk remained 
very hot, as evidenced by an extremely blue continuum and the presence of
high excitation lines.

IUE observations in 1987 (Krautter \& Snijders 1990) provided a measurement of
E(B-V) = 0.55 from the 2200\AA\ feature and a subsequent distance near 1200
pc. A later IUE observation obtained in 1991 (Gonzalez-Riestra et al., 1998) 
revealed the spectrum of this nova still showed emission lines of CIII (1175), SiIV (1400), NIV], 
CIV (1550), HeII (1640) but no CIII] (1909). The UV energy distribution became 
increasingly flatter during the time period when IUE spectra were obtained (1987-1991)
, indicating the gradual cooling of the
 heated white dwarf. On the basis of a 
non-detection of V842 Cen during the ROSAT survey, Gonzalez-Riestra et al.
(1998) concluded that nuclear burning had largely shut off, thus giving a 
turn-off time of approximately 3.5 years.

Two different studies have provided evidence that the 
system is being viewed at low orbital inclination, Woudt \& Warner (2003) from
the lack of orbital variation and 
Schmidtobreick et al. (2005) from the width of the Balmer lines. Schmidtobreick et al. (2005) interpret their 
spectra and photometric observations as suggesting that V842 Cen is a low
mass transfer object 
intermediate between novae and dwarf novae. This in turn suggests that
V842 cen could represent an important test case for the hibernation theory of 
cataclysmic variables (Shara et al. 1986).

High speed photometry obtained in 2000 (Woudt \&
Warner 2003)
showed no persistent short timescale periodicity, but further photometry in
2008 
(Woudt et al. 2009) revealed light 
curves that exhibited a coherent light modulation at 56.825 s, which they 
took as the rotation period of the white dwarf. Sidebands were apparent that
were indicative of 
re-processing from a surface moving with an orbital period of 3.94 h. Based 
upon their observations, they classified V842 Cen as an intermediate polar (IP) 
of the DQ Herculis subclass (Woudt et al. 2009). If the 57 second periodicity is the rotation period
 of the accreting white dwarf, then V842 Cen would contain the 3rd most rapidly 
rotating white dwarf among IPs. The fact that a 57s periodicity was seen suggests that the mass transfer rate had declined significantly since the nova outburst.

Luna et al. (2012) obtained XMM-Newton X-ray spectra in 2011 which were consistent with 
the emission from an absorbed thin thermal plasma with a temperature distribution
given by an isobaric cooling flow. From their cooling flow model fit and its corresponding $kT_{max}$ (43 keV), they derived a very high temperature in the shocked region. Following the method of Yuasa et al. (2010), they obtained a mass for the white dwarf in V842 Cen of $M_{wd}= 0.88 M_{\odot}$. Surprisingly however, they 
failed to detect the 57s periodicity found by Woudt et al. (2009), in either 
the OM data or the EPIC data. In addition, the X-ray luminosity appears to
be much lower than other IPs, especially considering the system's low
inclination and thus direct visibility of the accretion region.

To make further progress in understanding the decline and return to quiescence 
of this post-nova as well as the physical properties of the hot primary 
component, we obtained a Hubble Space Telescope (HST) ultraviolet spectrum along with
ground-based optical and near-IR photometry. These observations, an 
analysis of the UV spectrum with synthetic spectra, and our conclusions are presented in the sections below.

\section{Hubble Space Telescope COS Observation}

We observed V842 Cen with HST on 18 March 2010
from 16:50:05 to 17:27:59 UT. The instrumental setup used the COS/FUV configuration through the primary science aperture with the G140L grating in TIMETAG mode for an exposure time of 2,274 seconds.There were no reported problems during 
the observation and the obtained spectrum had a S/N$\sim$20:1. The data were downloaded from MAST and processed through the standard pipeline. 
The default extraction for the G140L grating is 57 pixels.
We de-reddened the data by adopting a color excess E(B-V) = 0.55 and using the IDL routine UNRED. The original COS spectrum, smoothed but without de-reddening, is displayed in Fig.1. In Fig.2, we display the COS spectrum of V842 Cen, de-reddened with E(B-V) = 0.55 with the identified line features labelled and artifacts identified in the figure caption. In order to optimize the spectrum for the light curves, we extracted with the optimum width of 41 pixels as found by maximizing the S/N through extractions using a series of widths.

\section{Optical and Near-IR Photometry}

Optical time-series data were obtained using the 74-in Radcliffe telescope
at the Sutherland site of the South African Astronomical Observatory with
the University of Cape Town CCD and no filter. Data were obtained from
1:28:48-2:12:41 UT on March 18. Optical V and J magnitudes were also
obtained with the 1.3m telescope of the 
Small and Moderate Aperture Research Telescope System on Cerro Tololo on the night of March 18 at 7:40 UT. 
Calibration of stars
in the field with the AAVSO chart sequence gave a V magnitude of 16.5 on
this date, while the 2MASS stars provided a J magnitude of 15.5.

\section{UV and Optical Light Curves and Periods}

A light curve was created from the COS spectrum by binning the time-tag data into 6 sec bins (comparable to the optical data) and summing the fluxes over all useful wavelengths from 1127-2000\AA. The resulting lightcurve was converted to fractional intensity for DFT analysis. 
The optical light curve was treated in the same manner. Then the DFTs were computed and the white noise determined empirically with the shuffling technique described in Kepler (1993) and Mukadam et al.(2010). 

The top panel of Figure 3 shows the original COS light curve and DFT. Since
there was a large change in flux from the start to the end of the observation, 
the trend
was removed and the resulting residual light curve is shown in the second
panel. The computed DFTs of the original and residual light curves are shown 
in the two bottom panels. The dashed line in the residual DFT shows the 
3$\sigma$
limit for the white noise as 9.5 mma (with unit of milli-modulation
amplitude). There is no visible short period
evident near this level.

The optical light curve was treated in the same fashion and 
is shown in Figure 4. The length of the optical light curve is
similar to the COS data, but was obtained about 15 hours earlier.
For the optical data, we used a standard IRAF (Tody 1993) reduction to 
extract sky-subtracted light curves from the CCD frames using PSF fitting. 
After a preliminary reduction, we converted the lightcurve to differential 
magnitude and the mid-exposure times of the CCD images to heliocentric 
julian date (HJD). Magnitudes are used for
the original light curve (top panel) to provide an easy comparison to past 
data of Woudt et al. (2009). As with the UV data, the decreasing long trend in
the light curve was removed to create a residual light curve (in fractional
intensity) in the second panel. The bottom panels show the resulting DFTs.
There is clearly a peak at 56.5$\pm$0.3s with an amplitude of
8.9$\pm$4.2 mma. While the noise level is high due to the short duration
of the light curve, this period is greater than the 3$\sigma$ noise level
of 7.5 and consistent with the period present in the 2008 data of
Woudt et al. (2009). The amplitude is higher than the value in 2008 (4.2mma).

\section{Synthetic Spectra of Photospheres and Disks}

We Constructed a grid of theoretical, high gravity, solar composition photospheric spectra
by first using the code TLUSTY (Version 203) (Hubeny 1988) to calculate the 
atmospheric structure which served as input to build the synthetic spectra with the 
code SYNSPEC (Version 48) (Hubeny and Lanz 1995). We compiled a library of photospheric spectra covering the 
temperature range from 15,000K to 70,000K in increments of 1000 K, and a 
surface gravity range, log $g = 7.0 - 9.0$, in increments of 0.2 in log $g$ (Ballouz and Sion 2009).

We also used model accretion disks extracted from the optically thick, accretion disk model library  of 
Wade \& Hubeny (1998). For these models, we took the innermost disk radius,
R$_{in}$, to be fixed at a fractional white dwarf radius
of $x = R_{in}/R_{wd} = 1.05$. The outermost disk radius, R$_{out}$, was selected such that 
T$_{eff}(R_{out})$ is 10,000K since disk annuli beyond this radius have cooler zones with larger radii, thus providing only a very small 
contribution to the mid and far UV disk flux, including the COS FUV bandpass.
The disk models are steady state in the sense that the mass accretion rate is assumed to be the same for all radii. With these definitions, the run of
disk temperature with radius is expressed as:

\begin{equation}
T_{eff}(r)= T_{s}x^{-3/4} (1 - x^{-1/2})^{1/4}
\end{equation}

where  $x = r/R_{wd}$
and $\sigma T_{s}^{4} =  3 G M_{wd}\dot{M}/8\pi R_{wd}^{3}$

Limb darkening of the disk is fully taken into account in the manner described 
by Diaz et al. (1996) involving the Eddington-Barbier relation, the increase of 
kinetic temperature with depth in the disk, and the wavelength and temperature 
dependence of the Planck function (Ballouz \& Sion 2009). The boundary layer contribution is ignored in our model fitting. Thus, the boundary layer is expected to be important only in the extreme ultraviolet below the Lyman limit. Disk truncation was not included in this work. 

In addition, for the purpose of handling disk truncation, we also 
constructed standard alpha accretion disk models for which we could
vary the disk truncation radius. These models, like the Wade and Hubeny (1998) grid
are standard alpha disk models (Shakura \& Sunyaev 1973), having the viscosity due to 
MHD turbulence, and with energy dissipated locally in the vertical direction. Each disk model consisted
of N rings of radii $r_i$ (i = 1, 2, ..N), where the temperature $T(r_i)$ and density ($\rho(r_i)$) of each ring are given by the standard disk model.
The codes TLUSTY \& SYNSPEC are then used to generate spectra for each of the disk rings, with the white dwarf mass, mass accretion rate, and radius of the ring as input. The individual spectra of the rings are then added together to generate a disk spectrum for a given inclination and assumed limb darkening (Ballouz and Sion 2009; Godon et al.2012).

Our fitting procedure is as follows.
The best-fitting 
single temperature white dwarf model and the best-fitting accretion disk-only 
model to the observed spectrum were found. Depending upon the success of these fits, we assessed the possibility of whether a combination
of a disk plus a white dwarf resulted in a statistically
significant improvement to the fits compared iwth that achieved with disks or photospheres
alone. Using two $\chi^{2}$ minimization routines, either IUEFIT for disks-alone and 
photospheres-alone or DISKFIT for combining disks and photospheres or 
two-temperature white dwarfs, $\chi^{2}$ values and a scale factor were 
computed for each model or combination of models (Godon et al.2012). The scale factor, $S$, normalized to a kiloparsec and solar radius, can be related to the white dwarf radius R through: 
$F_{\lambda(obs)} = S H_{\lambda(model)}$, where $S=4\pi R^2 d^{-2}$, and $d$ is
the distance to the source. For the white dwarf radius, we use the mass-radius 
relation from the evolutionary model grid of Wood (1995) for C-O cores. The 
best-fitting model or combination of models was chosen based not only upon the 
minimum $\chi^{2}$ value achieved, but also the goodness of fit of the continuum 
slope, the goodness of fit to the observed Lyman Alpha region and consistency of
the scale factor-derived distance with other distance estimates (Ballouz et al.2009).

\section{Fitting Strategy and Results}

Given that the V842 Cen is a relatively young
post-nova, it was expected that the white dwarf 
remains very hot and that the accretion rate should be quite high due to irradiation and/or bloating
of the donor star by the nova event. However, the spectroscopic study by
Schmidtobreick et al.(2005) presented evidence that the accretion rate in V842 Cen is 
considerably lower than expected. Moreover, the XMM study of Luna et al.(2012) revealed a surprisingly low accretion rate, given that the system underwent a nova so relatively recently. Indeed, the X-ray luminosity of V842 Cen appears to be below the flux level of other low inclination IPs where the accretion region is directly visible (Woudt et al.2009). If the accretion rate is as low as indicated by Luna et al. (2012), then it is possible that
the underlying white dwarf photosphere is exposed in this old nova since the disk luminosity and the temperatures/luminosities of the accretion spots would be low enough that the white dwarf photospheric FUV flux dominates.
However, the problem with seeing primarily the white dwarf or a low disk luminosity is that the V magnitude of the system is still  2 magnitudes brighter than at quiescence.  An argument in favor of a disk is that the V-J color of 1 is more consistent with a disk than a hot WD. If the underlying WD photospheric in V842 Cen is exposed, then this is a rarity among IPs since they are not known to undergo low states and the WD photospheric light is typically swamped by the accretion light of the heated polar spots and the accretion curtains.

Given that the intermediate polar status of V842 Cen is not solidly confirmed but rests solely on the optical data, we explored model fitting using full  accretion disk models alone, as though the WD is non-magnetic. Since the low accretion rate in this old nova suggests that the white dwarf is exposed, then we also explored model fitting of the HST COS data with single temperature white dwarf photospheres.
 
Our first fitting experiment implicitly assumed that V842 Cen is a non-magnetic nova. Therefore, we utilized single temperature, solar composition white dwarf models. For this procedure, we fitted the COS spectrum with a total of 560 photosphere models between 15,000K and 70,000K with log g varying from 7 to 9. Without exception, the fits were poor with either the Lyman alpha profile being far too broad when the continuum had the correct slope or with fair agreement to the Lyman Alpha profile but a markedly incorrect continuum slope. Moreover, the distances yielded by the photosphere fits are implausibly close. Hence, we have ruled out that the COS spectrum arises from a white dwarf photosphere. In Figure 5, a typical photosphere fit to the COS spectrum is shown with Teff = 31,000K, Log g = 8.6.

Our next fitting attempt used full accretion disks from the model grid of Wade and Hubeny (1998). There is substantial observational evidence supporting a low inclination for V842 Cen but we explored all inclinations in the grid models. Out of the roughly 900 disk models using every combination of $i$, M$_{wd}$, we tried to pinpoint the best-fitting
accretion-disk model, WD model or combination of a disk and WD.
The full disk models yield quite good fits for low mass accretion rates, leading to distances of a few hundred parsecs. The inclusion of a model WD with the disk model fits did not yield any significant improvement.
The best fitting full accretion disk fit has 
$\dot{M} = 10^{-10} M_{\odot}$ yr$^{-1}$, $i = 41 \degr$, but a distance 
of only 210 pc if the WD mass is $0.8 M_{\odot}$. 
If $M_{wd}= 1.2 M_{\odot}$, then d = 281 pc. To obtain greater distances, close to 1 kpc, then Mdot has to be > \.{M}$ = 3.0\times 10^{-9}$ M$_{\odot}$ yr$^{-1}$, for $M_{wd}= 0.8-1.2 M_{\odot}$. All lower mass white dwarfs (hence larger radii) are ruled out as the distance becomes shorter. In Figure 6, we display the best-fitting full disk. The blue regions in the figure are masked out of the fit for which the $\chi^2 = 5$. Lower and higher inclinations than $i = 41 \degr$ do not yield a distance $> $250 pc. Models that yielded larger distances up to 500pc deviate significantly and have much larger $\chi^2$ values.  

Given the strong possiblity that V842 Cen is an intermediate polar, 
trunacted accretion disk models are needed for comparison with the HST COS data. But for a magnetic accretor, additional free parameters are introduced to the fitting such as the field strength, magnetic moment, disk truncation radius, and flux components associated with the accretion spots and the accretion curtains. Since the magnetic field strength of the white dwarf is unknown, the truncation radius of the accretion disk is a free parameter. We made a first guess at the magnetic moment of the white dwarf by comparing it with magnetic WDs in 
intermediate polars tabulated by Belle \& Sion (2009) and Mukai (2012). We created full accretion disk models using TLUSTY and SYNSPEC. We then subtracted out inner annuli that had radii less than the inner disk radius.  Because accretion disks do not have UV flux-contributing annuli past ~ $R_{wd}$, we used the inner disk radius calculated from the optical emission line wings. From optical emission line wing velocities, assuming a Keplerian 
disk, $R_{in}$ = 2.5 - 9.0 $R_{wd}$. We then subtracted out inner annuli that had radii less than the inner disk radius. We used the magnetic parameters of EX Hydrae as approximate input for our truncated disk models for V832 Cen since the WD masses and lower accretion rates of the two systems appear similar. We adopted a magnetic 
moment of  $\sim 5 \times 10^{33}$ Gcm$^3$ 
(Norton et al.2004) and truncation radii of 3 $R_{wd}$ and 5 $R_{wd}$.

We carried out truncated disk fits to the COS data for truncation radii of 3 $R_{wd}$ and 5 $R_{wd}$. 
For both truncation radii, and for disk inclinations of $i = 15 \degr$, $i = 40 \degr$, and $i = 60 \degr$, none of the fits were 
satisfactory. For $M_{wd}= 0.8-1.2 M_{\odot}$, the entire disk is only 10 
$R_{wd}$ to 12 $R_{wd}$ in diameter so a truncation of the disk radius 
in that range leaves only a few annuli or no annuli at all. 
In Figure 7, we display a truncated disk fit for a disk inclination
$i = 15 \degr$, a WD mass $M_{wd} = 0.8 M_{\odot}$ and accretion rate 
$\dot{M} = 3.0\times 10^{-10}$ M$_{\odot}$ yr$^{-1}$. 
Figures 8 and 9 show 
truncated model fits for the same parameters as Figure 7 but with disk 
inclinations of $i = 40 \degr$ and $i = 60 \degr$, respectively. 
If the accretion rate were increased above 
$\dot{M} = 10^{-10}$ M$_{\odot}$ yr$^{-1}$, then the accretion rate and diameter of the truncated disk would increase but the increased mass residing in the outer disk and the higher accretion rate would contradict the observations of Schmidtobreick et al.(2005), Woudt et al. (2009), and Luna et al.(2012) demonstrating that the accretion rate is low.
  
\section{Discussion}

The optical spectroscopic and photometric studies, including the optical data presented in this
work, lend support to the classification of V842 Cen as an intermediate polar. However, there are major difficulties with the intermediate polar classification. The lack of a detection
of the spin period in the UV COS data is also a problem.
The non-detection of X-rays by three different X-ray observatories, ROSAT, SWIFT and XMM-Newton and hence no X-ray signature of spin casts serious doubt on the IP classification. Moreover, our best-fitting, and simplest, accretion disk model of the HST COS observation is a full optically thick, steady state accretion disk with no magnetic truncation and a distance shorter than 500 pc. Our model fitting of the HST COS data rules out a white dwarf photosphere as the source of the FUV flux and indicates a surprisingly low accretion rate for a classical nova only 26 years after the explosion in 1986 when the accretion rate is expected to be high and the white dwarf should still be very hot. Indeed, the best model fit that we achieved was with a full optically thick disk appropriate for a non-magnetic white dwarf. 
This best-fit model has a solar mass white dwarf, $i = 40 \degr$, \.{M}$ = 3.0\times 10^{-10}$ M$_{\odot}$ yr$^{-1}$ with a scale-factor derived distance of several hundred pc but still well below the distance of Sekiguchi et al.(1989) and Gill and O'Brien (1998) of ~1 kpc.

Despite the indicators discussed above, one cannot rule out that V842 Cen is an IP, based upon the optical observations alone. However, definitive proof of an IP usually requires confirmatory X-ray observations (K. Mukai, private communication). It is possible that the X-ray emission was either gone or reduced in amplitude at the time of the X-ray observations or that the X-rays were absorbed by material above the disk plane. It is possible that the magnetic field of the white dwarf in V842 Cen is weaker than in other IPs with stable rotational signals from rotating white dwarf primaries (Woudt et al. 2009). In this scenario, a more weakly magnetic rotating white dwarf would not exhibit a strong rotational signal for very long after the nova event because the magnetosphere of the WD is crushed by the ram pressure of the accretion flow. On the other hand, this scenario for V842 Cen requires that the enhanced accretion flow induced by irradiation of the donor be unexpectedly short-lived and the cooling of the very hot, post-nova white dwarf would be surprisingly rapid.
The observational constraint that V842 Cen's supersoft phase was no longer than 3.5 years may support extraordinarly rapid post-outburst evolution. In any event, the accretion rate is surprisingly low and the hot component cooler than expected for only 24 years post-nova, especially if irradiation of the donor star took place. Our interpretation remains fraught with puzzles. Why is the spin not evident in the FUV (or X-rays)? Why is the V-magnitude two magnitudes brighter than pre-outburst if the accretion rate is low? Why is there a large variation (in UV flux (and optical to a lesser degree) during the 40 minutes of observation when, at low inclination,a full disk should be flat? Clearly, further observations and modeling are required before this conclusion becomes irrefutable.
The fact is that our present knowledge of post-nova behavior is poor and the broad spectroscopic diversity seen among post-novae with similar elapsed times since their outbursts is not yet understood.   

Acknowledgements

We thank Ramotholo Sefako and Peter Lake for contributing optical
magnitudes. We are grateful for the support of HST grant GO-11639 and NSF grant AST-1008734.

References 

%\bibitem[]{}
Ballouz, R., \& Sion, E.2009, ApJ, 697, 1717                    

%\bibitem[]{}
Belle et al. 2003, ApJ, 587, 373

%\bibitem[]{}
Belle, K., \& Sion, E.2009, BAAS, 41, 469 

%\bibitem[]{}
Diaz, M., Wade, R., \& Hubeny, I., 1996, ApJ, 459, 236

%\bibitem[]{}
Gill C.D., O'Brien, T.J., 1998, MNRAS, 300, 221

%\bibitem[]{}
Godon, P., Sion, E.M., Levay, K,. Linnell, A.P., Szkody, P., Barrett, P., Hubeny, I., Blair, W.2012, ApJS, 203, 29

%\bibitem[]{}
Gonzalez-Riestra,R., Orio, M., \& Gallagher, J.1998, A\&A Suppl.,129, 23

%\bibitem[]{}
Hubeny, I. 1988, Comput. Phys. Comun., 52, 103

%\bibitem[]{}
Hubeny,I., \& Lanz, T.1995,ApJ, 439, 875

%\bibitem[]{} 
Kepler, S.~O.\ 1993, Baltic Astronomy, 2, 515

%\bibitem[]{}
Krautter, J. \& Snijders, M.A.J. 1990 in Cataclysmic Variables and
Low Mass X-ray Binaries, ed. C. W. Mauche, CUP, p. 387

%\bibitem[]{}
Luna, G., Diaz, M., Brickhouse, N., \& Morales, M.2012, MNRAS, in press

%\bibitem[]{}
McNaught, R.H. 1986, IAUC 4274

%\bibitem[]{}
Mukadam, A. S. et al. 2010, ApJ, 714, 1702

%\bibitem[]{}
Mukai, K.2012, Intermediate Polar Catalog, Version 2011a 

%\bibitem[]{}
Norton et al. 2004, ApJ, 614, 349; 

%\bibitem[]{}
Pandel et al., 2005, ApJ, 626, 396

%\bibitem[]{}
Schmidtobreick,L., Tappert,C., Bianchini, A., \& Mennickent, R.2005, A\&A, 432, 199

%\bibitem[]{}
Sekiguchi K., Feast M. W., Fairall A. P., \& Winkler H., 1989,MNRAS, 241, 311

%\bibitem[]{}
Shara, M.M., Livio, M., Moffat, A.F.J., \& Orio, M. 1986, ApJ 311, 163

%\bibitem[]{} 
Tody, D.\ 1993, Astronomical Data Analysis Software and Systems II, 52, 173

%\bibitem[]{}
Wade, R., \& Hubeny, I.1998, ApJ, 

%\bibitem[]{}
Whitelock, P. 1987, MNSSA 46, 72

%\bibitem[]{}
Wood, M. 1995, in White Dwarfs, Proc. of the 9th European Workshop on White Dwarfs
Held at Kiel, Germany, 29 Aug.1 Sep. 1994, Lecture Notes in Physics Vol. 443, ed.
D. Koester \& K. Werner (Berlin: Springer), 41

%\bibitem[]{}
Woudt P. A. \& Warner B., 2003, MNRAS, 340, 1011

%\bibitem[]{}
Woudt P. A., Warner B., Osborne J., \& Page K.,  2009, MNRAS, 395, 2177

%\bibitem[]{}
Yuasa T. et al., 2010, A\&A,520,A25

\begin{figure}
%\epsscale{0.5}
\plotone{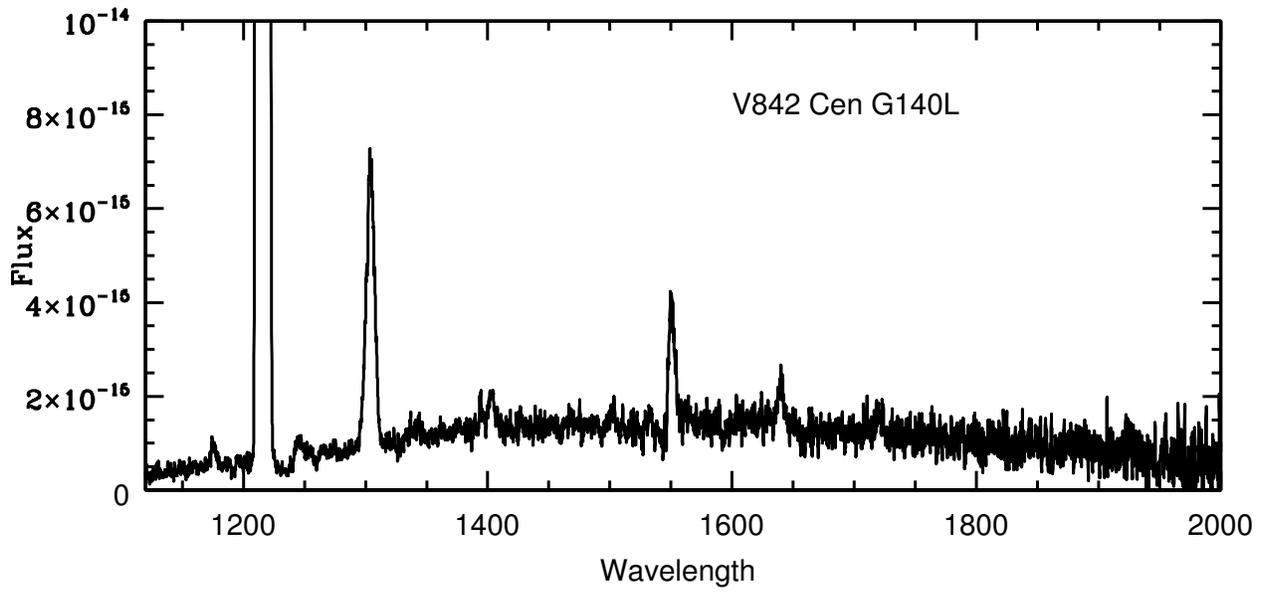}
\caption{The observed HST COS G140l spectrum of the old nova V842 Cen
before de-reddening. These data were
smoothed by 5 and then smoothed again by 3 (boxcar smoothing).}
\end{figure}

\begin{figure}
\begin{center}
\plotone{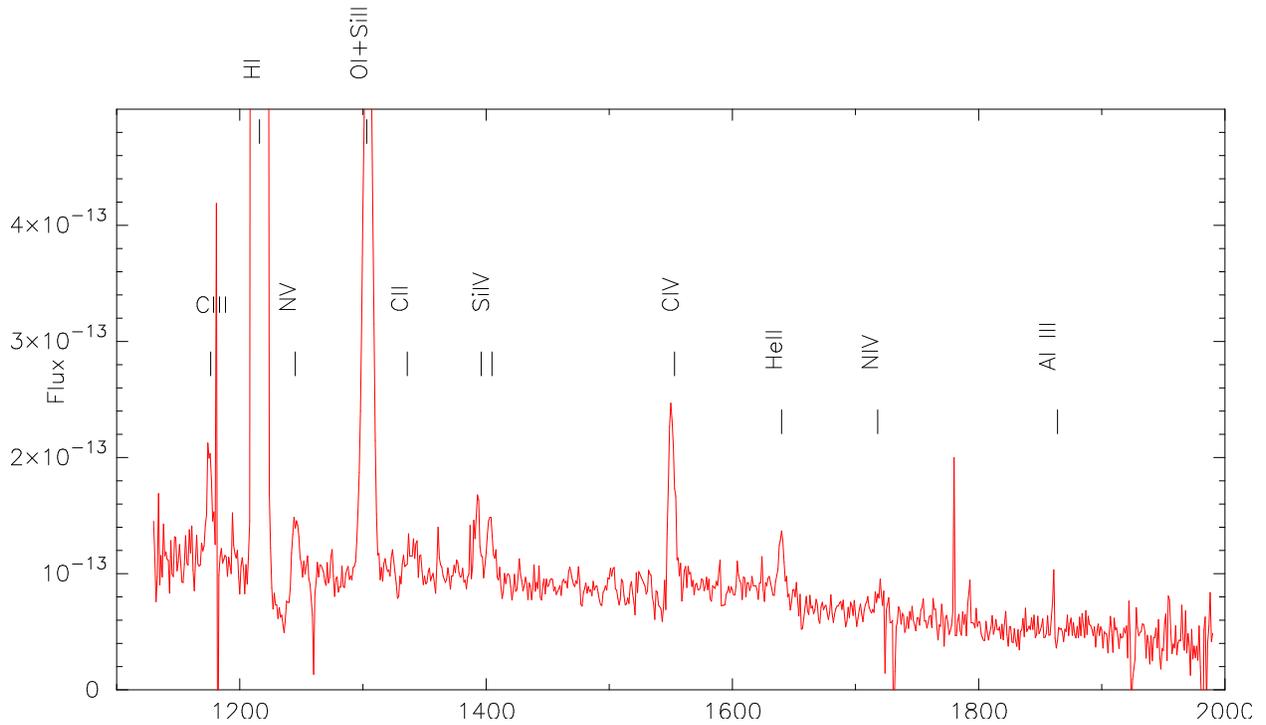}
\end{center}
\caption{{The observed HST COS G140l spectrum of the old nova V842 Cen
de-reddened with E(B-V) = 0.55 and with the most prominent spectral features identified. The data were downloaded from MAST and processsed through the standard pipeline. The sharp dips between 1256-1260 and 1723-1728 as well as the sharp spikes just longward of C III 1175 and near 1780, are all artifacts. The strong emission features at Lyman Alpha and O I + SiIII (1300) are primarily geocoronal in origin.}}
\end{figure}

\clearpage
\begin{figure}
%\epsscale{0.5}
\plotone{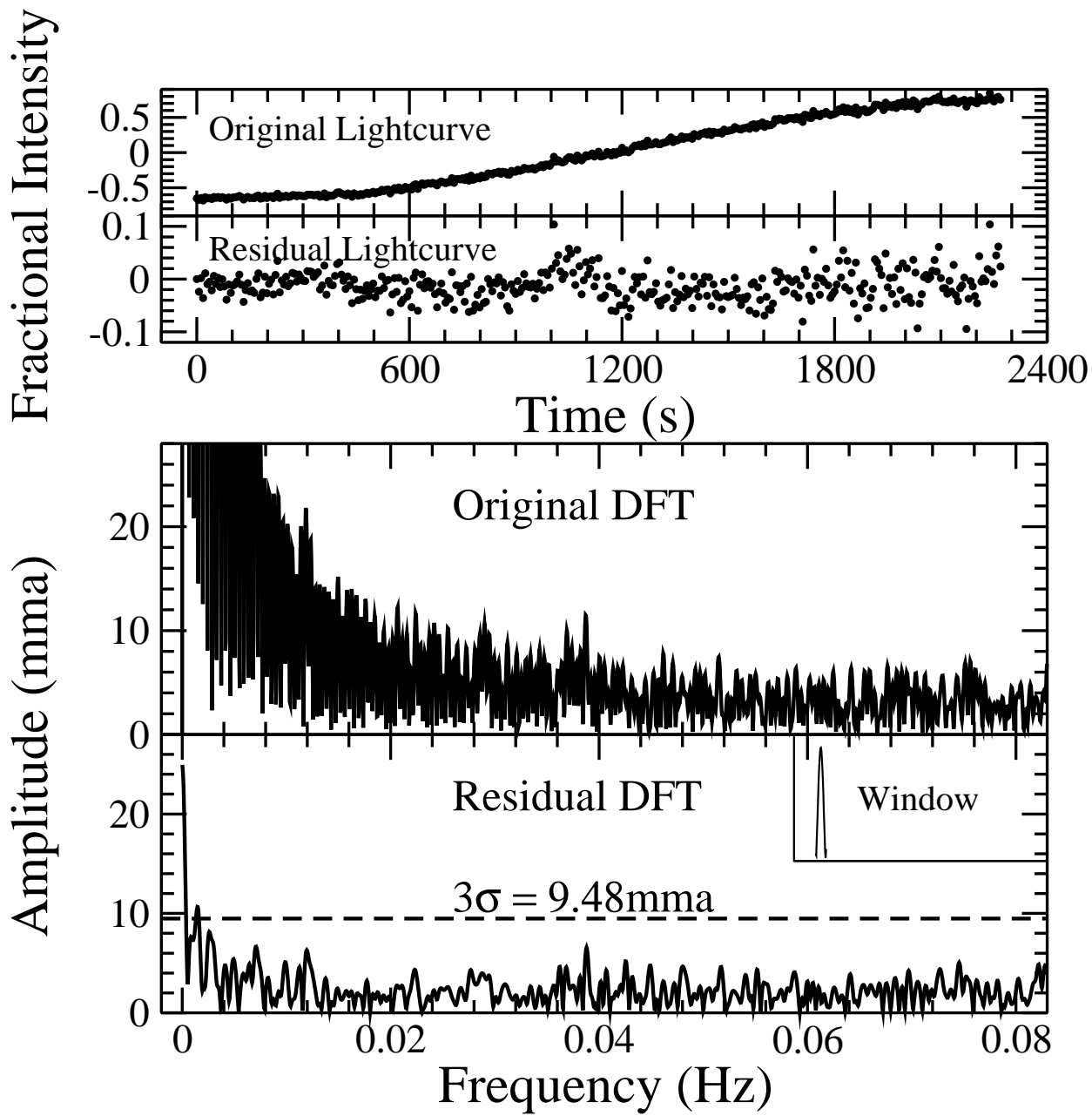}
\vspace{1.cm} 
\caption{top panel shows the original light curve created from the COS Time-Tag data, and DFT. Since
there was a large change in flux from the start to the end of the observation, 
the trend was removed and the resulting residual light curve is shown in the second
panel. The computed DFTs of the original and residual light curves are shown 
in the two bottom panels. The dashed line in the residual DFT shows the 
3$\sigma$limit for the white noise as 9.5 mma (with unit of milli-modulation
amplitude).}
\end{figure}

\clearpage
\begin{figure}
%\epsscale{0.5}
\plotone{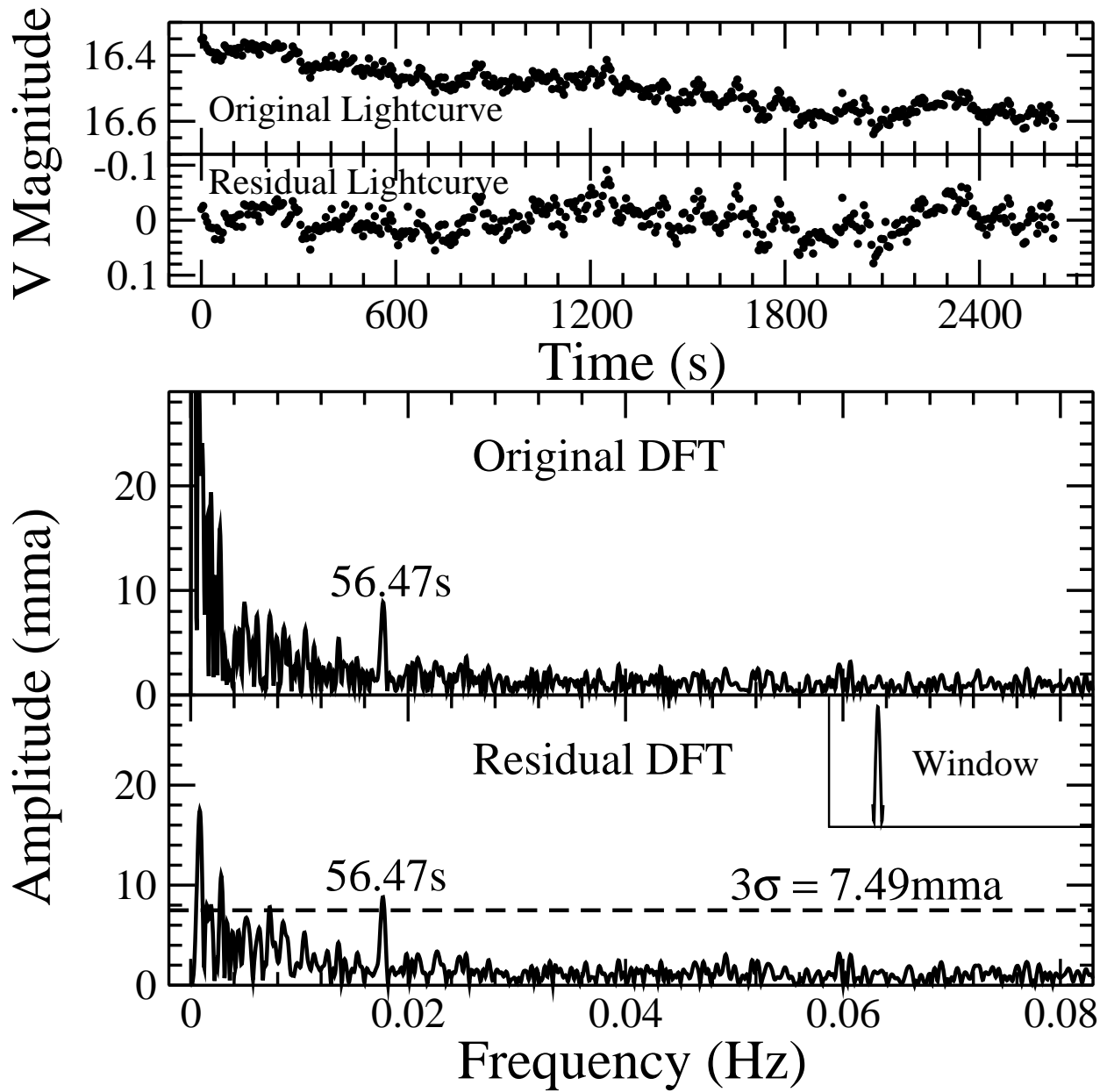}
\vspace{2.cm} 
\caption{ The optical light curve (treated in the same fashion as the COS light curve in Figure 3). 
The length of the optical light curve is
similar to the COS data, but obtained 15 hours earlier. 
In the top panel, magnitudes are used for
the original light curve data. The decreasing long trend in
the light curve was removed to create a residual light curve (in fractional
intensity) in the second panel. The bottom panels show the resulting DFTs.}
\end{figure}

\clearpage
\begin{figure}
%\epsscale{0.5}
\plotone{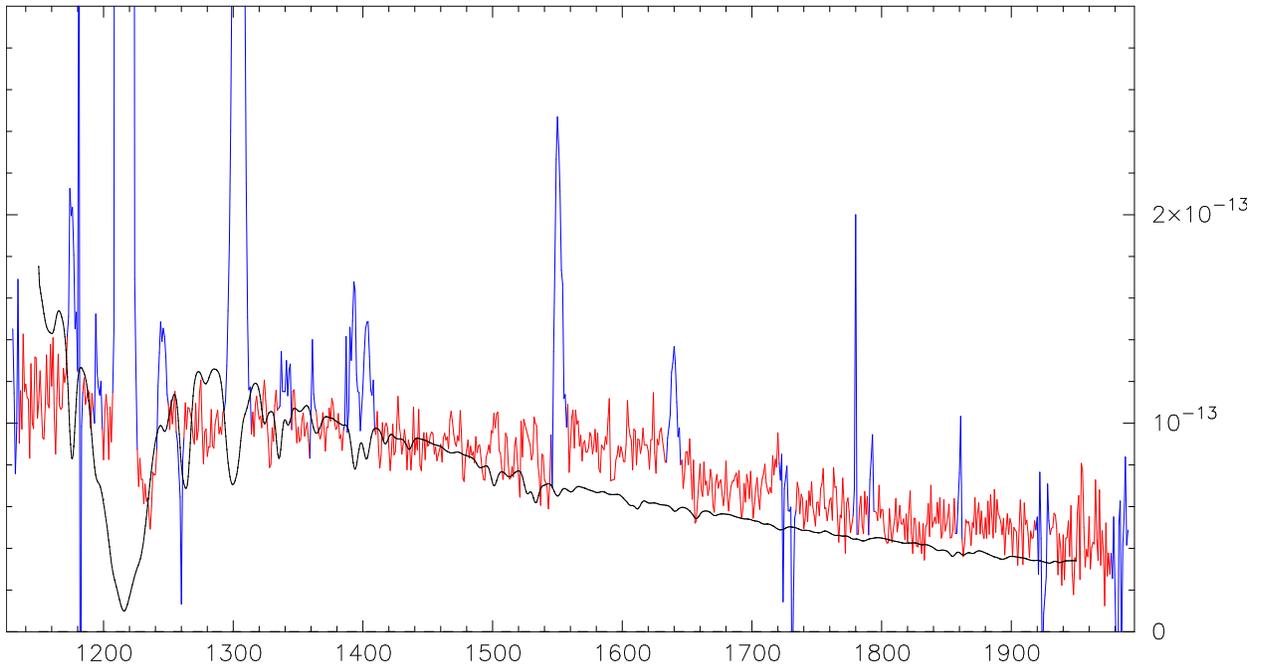}
\caption{The best-fitting white dwarf photosphere model to the HST COS spectrum 
of the old nova V842 Cen. The fit is implausible due to the short distance
it implies}
\end{figure}

\clearpage
\begin{figure}
%\epsscale{0.5}
%\plotone{v842fig6.ps}
\includegraphics[scale=0.60,angle=0]{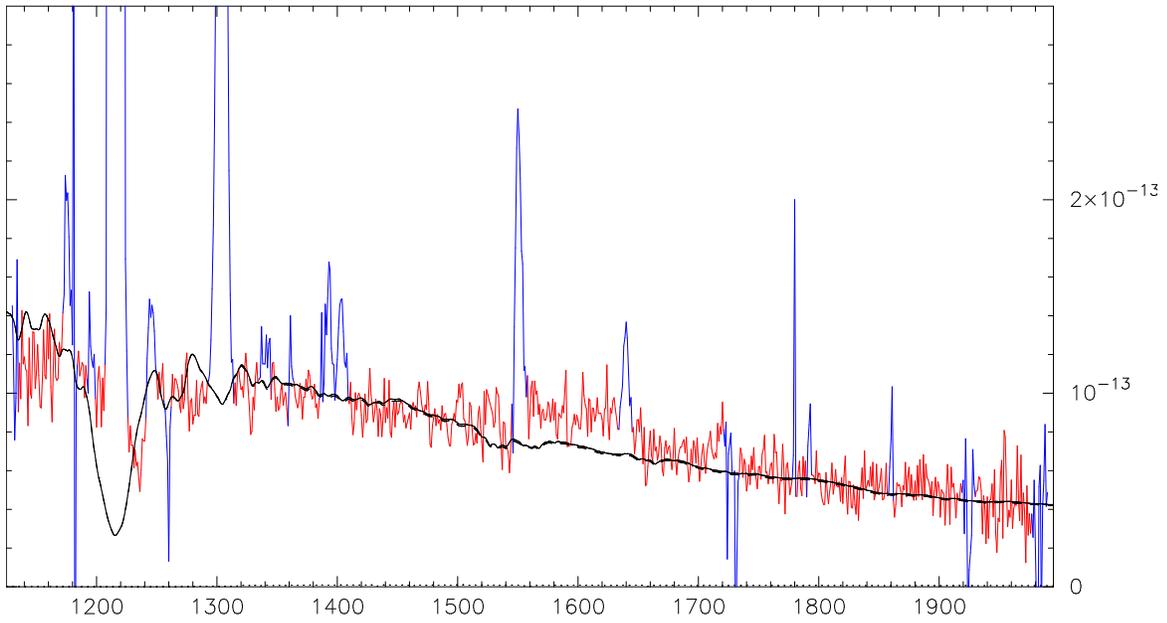}
\caption{The best-fitting full accretion disk model to the HST COS spectrum of the old nova V842 Cen taken 24 years after the nova explosion. 
The accretion disk corresponds to 
$\dot{M} = 3.0 \times 10^{-10} M_{\odot}$ yr$^{-1}$, see text for details.}
\end{figure}

\clearpage
\begin{figure}
%\epsscale{0.5}
\plotone{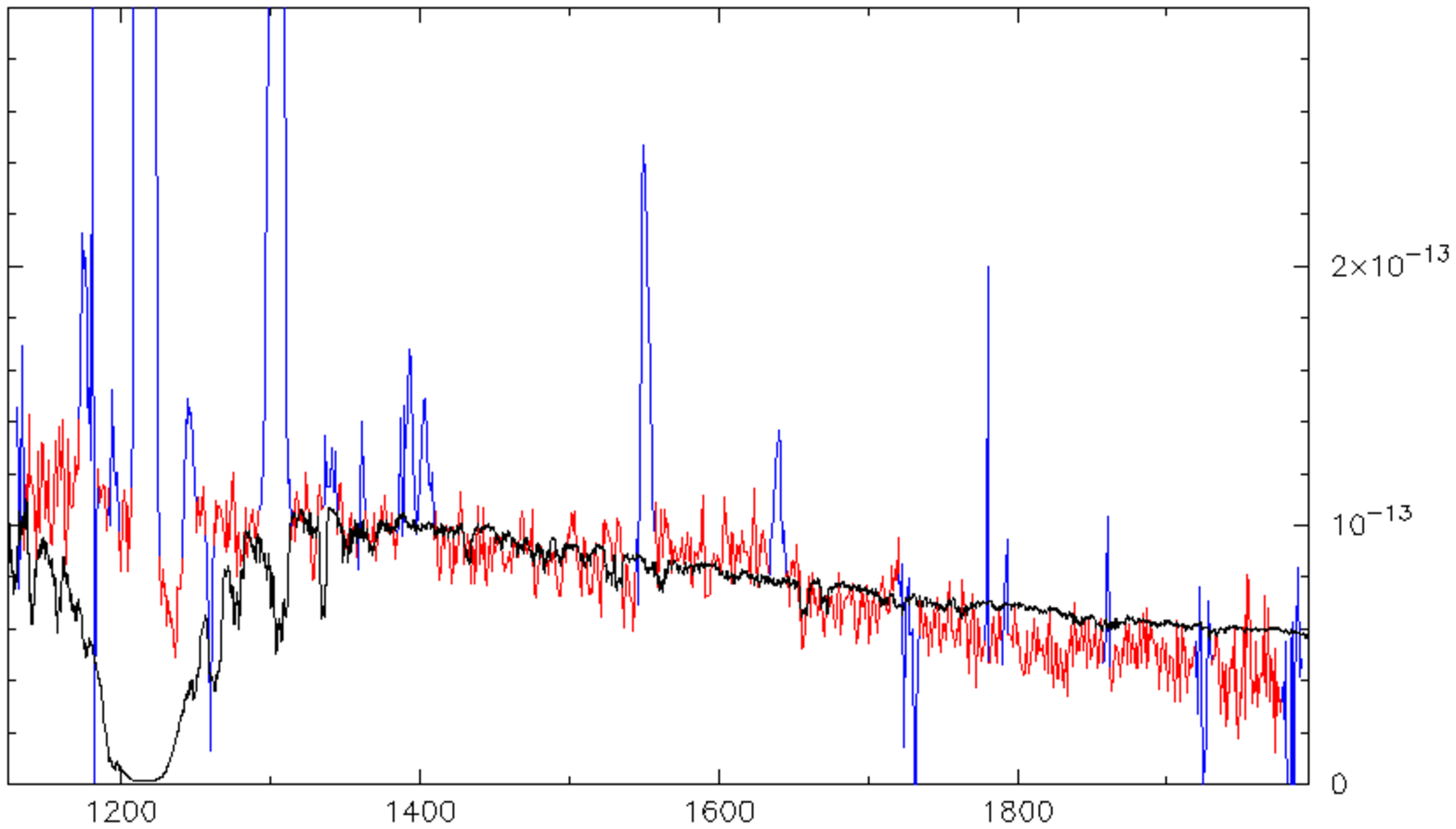}
\caption{
The best-fitting truncated accretion disk model to the 
HST COS spectrum of V842 Cen
For a disk truncation radius of $5 R_{wd}$ and inclination angle of 
$i = 15 \degr$.
The accretion disk corresponds to 
$\dot{M}= 3.0 \times 10^{-10} M_{\odot}$ yr$^{-1}$, see text for details.}
\end{figure}

\clearpage
\begin{figure}
%\epsscale{0.5}
\plotone{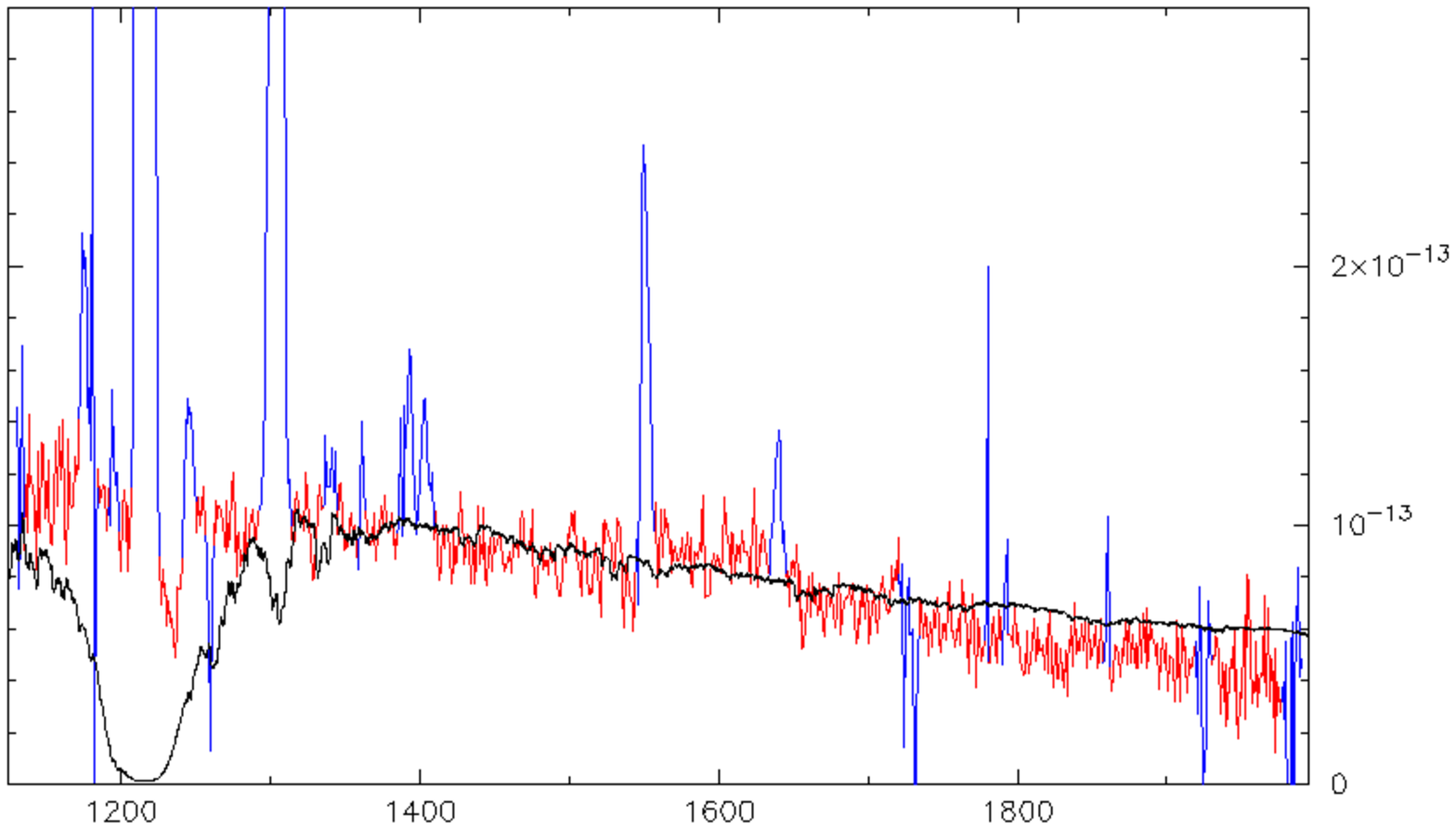}
\caption{The best-fitting truncated accretion disk model to the HST COS spectrum of V842 Cen
For a disk truncation radius of $5 R_{wd}$ 
and inclination angle of $i = 40 \degr$.
The accretion disk corresponds to 
$\dot{M} = 5.0 \times 10^{-10} M_{\odot}$ yr$^{-1}$, see text for details.}
\end{figure}

\clearpage
\begin{figure}
%\epsscale{0.5}
\plotone{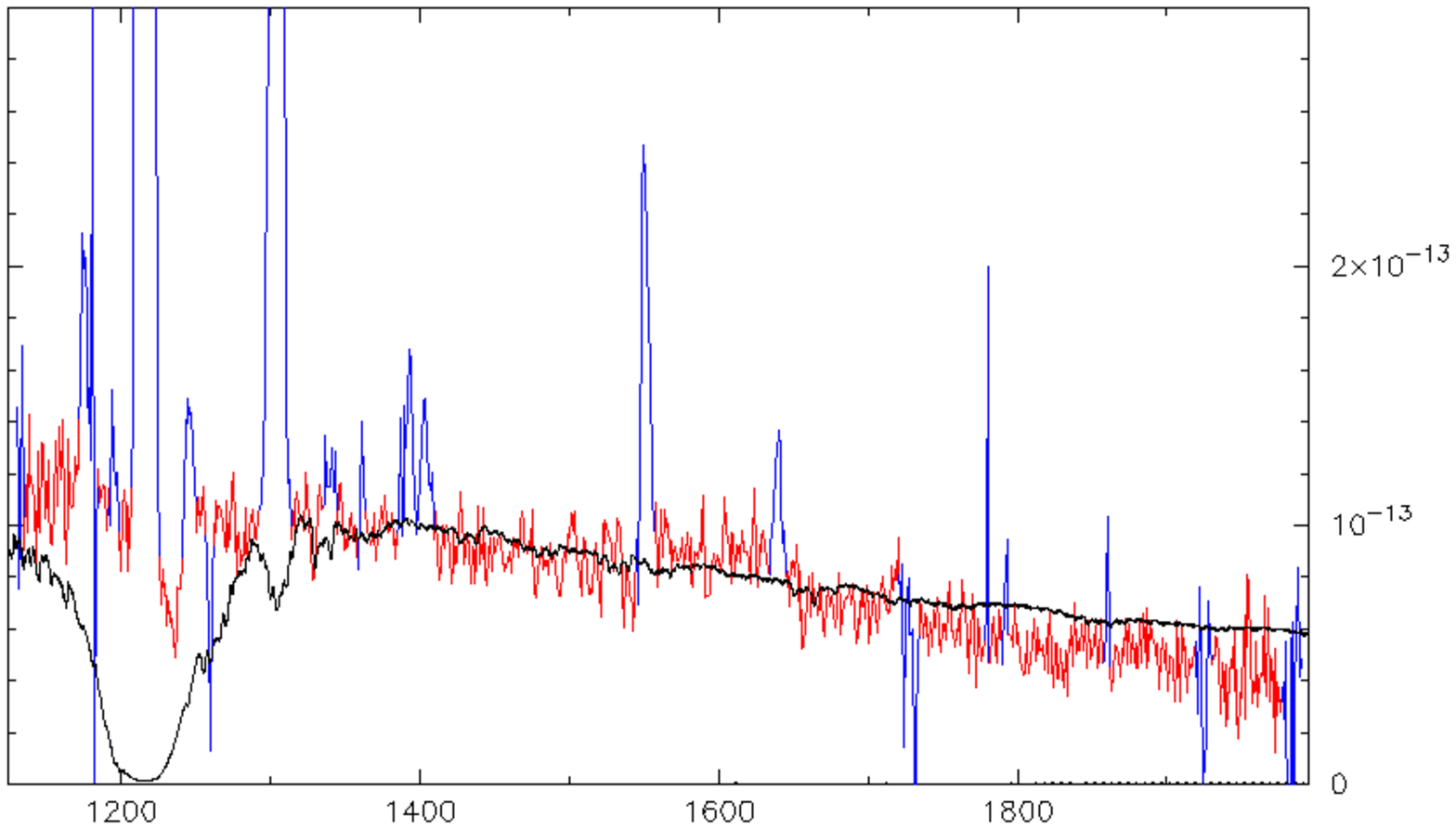}
\caption{The best-fitting truncated accretion disk model to the HST COS spectrum of V842 Cen
For a disk truncation radius of $5 R_{wd}$ and inclination angle of $i = 60 \degr$.
The accretion disk corresponds to $\dot{M} = 5.0\times 10^{-10} M_{\odot}$ yr$^{-1}$, see text for details.}
\end{figure}

\end{document}